\documentclass[floatfix,twocolumn,aps,prd,showpacs,amsmath,amssymb]{revtex4}

\usepackage{epsfig}
\usepackage{color}

\newcommand{\gam}{\gamma}

\newcommand{\sig}{\sigma}

\newcommand{\Aout}{A_\text{out}}
\newcommand{\Ain}{A_\text{in}}

\newcommand{\lmax}{l_m}
\newcommand{\sigabs}{\sigma_{\text{abs}}}
\newcommand{\sigscat}{\sigma_{\text{scat}}}
\newcommand{\sigtot}{\sigma_{\text{tot}}}

\begin{document}


 \title{Scattering of Sound Waves by a Canonical Acoustic Hole}

\author{Sam R. Dolan}
 \email{sam.dolan@ucd.ie}
 \affiliation{%
 School of Mathematical Sciences, University College Dublin, Belfield, Dublin 4, Ireland.
}%

\author{Ednilton S. Oliveira}
\email{ednilton@fma.if.usp.br}
\affiliation{Instituto de F\'\i sica, Universidade de S\~ao Paulo, 
CP 66318, 05315-970, S\~ao Paulo, SP, Brazil}

\author{Lu\'\i s C. B. Crispino}
\email{crispino@ufpa.br}
\affiliation{Faculdade de F\'\i sica, Universidade Federal do
Par\'a, 66075-110, Bel\'em, PA,  Brazil}

\date{\today}

\begin{abstract}
  This is a study of a monochromatic planar perturbation impinging
  upon a canonical acoustic hole. We show that acoustic hole
  scattering shares key features with black hole scattering. The
  interference of wavefronts passing in opposite senses around the
  hole creates regular oscillations in the scattered intensity. We
  examine this effect by applying a partial wave method to compute the
  differential scattering cross section for a range of
  incident wavelengths. We demonstrate the existence of a scattering peak
  in the backward direction, known as the glory. We show that
  the glory created by the canonical acoustic hole is
  approximately $170$ times less intense than the glory created by the
  Schwarzschild black hole, for equivalent horizon-to-wavelength
  ratios. We hope that direct experimental observations of such
  effects may be possible in the near future.
\end{abstract}

\maketitle

\section{\label{sec:introduction}Introduction}

Black holes play an important r\^ole in modern physics. For many years, they were considered merely a mathematical curiosity arising from Einstein's field equations. In more recent decades there has emerged strong evidence for their presence in binary systems and galactic centers. Aside from the growing observational data, it is clear that the \emph{idea} of a black hole has its own attractive power. Black holes have captured the imagination of a generation of physicists. For example, debates on black hole thermodynamics~\cite{Carlip-2007} and the information paradox~\cite{Giddings-Lippert-2004} are ongoing. The detection of gravitational waves from black hole binaries is hoped to be imminent~\cite{LIGO-2008}. The possibility of black hole creation at the LHC~\cite{Kanti-2004, Koch-Bleicher-Hossenfelder-2005} has recently received hyperbolic media attention.

Naturally, the possibility that \emph{black hole analogues}~\cite{Barcelo-Liberati-Visser-2005} may be created in the laboratory has attracted widespread interest. Analogues are artificial systems with some of the key properties of black holes. In 1981, Unruh~\cite{Unruh-1981} proposed the idea of a \emph{dumb hole}: a region of fluid from which no sound may escape. A dumb hole, or \emph{acoustic hole}~\cite{Visser-1998}, is bounded by an apparent horizon: a surface through which the normal flow velocity is equal to the speed of sound in the fluid. In this paper, we study the simplest spherically-symmetric dumb hole, the so-called \emph{canonical acoustic hole}. A range of alternative analogue systems have been proposed, for example in superfluid helium~\cite{Volovik-2003}, in Bose-Einstein condensates~\cite{Garay-2002}, in electromagnetic waveguides~\cite{Schutzhold-Unruh-2004}, in optical fibers~\cite{Corley-Jacobson-1999, Unruh-Schutzhold-2007} and other systems~\cite{Novello-Visser-Volovik}.

Under certain assumptions (an inviscid, barotropic fluid; irrotational flow), the Navier-Stokes equations describing small perturbations to fluid flow turn out to be formally identical to the equations for a massless scalar field propagating in a (3+1) Lorentzian geometry. That is~\cite{Visser-1998},
\begin{equation}
\Box \psi \equiv  \frac{1}{\sqrt{-g}} \partial_\mu \left( \sqrt{-g} g^{\mu \nu} \partial_\nu \psi \right) = 0 ,
\label{klein-gordon-eqn}
\end{equation}
where $\psi$ is a small perturbation to the potential describing the fluid velocity, $\mathbf{v} = \mathbf{v}_0 - \nabla \psi$. The effective metric $g_{\mu \nu}$ is an algebraic function of the background flow $\mathbf{v}_0$, the local density $\rho(P)$, and the speed of sound $c^{2} = \partial P / \partial \rho$, where $P$ is the fluid pressure. The appropriate metric for the canonical acoustic hole is given in section~\ref{sec-canonical}. 

Acoustic black holes are expected to emit a thermal spectrum of phonons, since Hawking's original arguments follow through without essential modification~\cite{Unruh-1981}. On the other hand, the laws of black hole thermodynamics do not follow. It seems that the association of entropy with horizon area is a special property of the Einstein equations~\cite{Visser-1998b}. The Hawking temperature for the canonical acoustic hole, $T_H \approx 1.2 \times 10^{-9} \, \text{K m} \, (c/1000\text{ms}^{-1}) (c^{-1} dv / dn)$~\cite{Visser-1998}, is sufficiently small that direct detection of Hawking radiation has not been achieved. The possibility of detecting Hawking emission from a 1D Bose-Einstein analogue is under investigation \cite{Carusotto-2008}.

There is more to acoustic holes than just Hawking radiation, however. As has been pointed out~\cite{Cardoso-2006}, there is a range of interesting classical wave phenomena which should be amenable to experiment. In particular, the scattering of perturbations by acoustic holes has been studied by a number of authors, in both the frequency domain~\cite{Basak-Majumdar-2003, Berti-Cardoso-Lemos-2004, Lepe-Saavedra-2005} and the time domain~\cite{Cherubini-Federici-Succi-Tosi-2005}. Like black holes, acoustic holes have characteristic damped resonances called quasi-normal modes. These resonances appear in the dynamic response of acoustic holes to external perturbation. Time-dependent scattering has been considered in detail for  the lowest multipoles $l=0,1, \ldots$. However, a study of the overall scattering pattern produced by the superposition of many angular modes has not been considered. This is the main purpose of our paper.

In this study, we adopt the partial wave approach to investigate the scattering and absorption of a monochromatic external perturbation impinging upon a canonical acoustic hole. We compare our results with previous studies of time-independent scattering by astrophysical holes~\cite{Futterman-1988}. In this regard, we are motivated by the exciting prospects for gravitational wave astronomy~\cite{LIGO-2008}. It is possible that acoustic scattering experiments may offer a way to better understand the dynamics of gravitational radiation. Further studies of wave propagation in analogue systems and their experimental realization are to be encouraged.

The monochromatic scattering scenario depends on only a single dimensionless parameter~\cite{Crispino-Oliveira-Matsas-2007},
\begin{equation}
\omega r_h = 2 \pi r_h / \lambda
\end{equation}
which expresses the ratio of the incident wavelength $\lambda$ to the horizon size $r_h$ of the canonical acoustic hole. For convenience, we have set the speed of sound $c$ equal to unity from here on.

The remainder of the paper is organized as follows. In section~\ref{sec-canonical} we introduce the canonical acoustic hole, and the equations which describe the propagation of sound perturbations in the effective geometry. In section~\ref{sec-orbits} we examine the properties of geodesics on a class of spherically-symmetric geometries. We derive strong and weak-field approximations for the scattering angle, and extend the glory approximation~\cite{Matzner-1985} to the case of the canonical acoustic hole. In section~\ref{sec-partialwave} we review the partial-wave method for time-independent scattering. An approximation for the phase shift in the large-$l$ limit is obtained. In section~\ref{sec-numerical} we discuss the details of the numerical methods employed. A selection of key results is presented in section~\ref{sec-results}. We conclude with our final remarks in section~\ref{sec-finalremarks}. The metric signature $(+---)$ is used throughout.

\section{The Canonical Acoustic Hole\label{sec-canonical}}
Let us consider a spherically-symmetric steady flow of incompressible fluid in three dimensions, with a source or sink at the origin $r=0$. 
Conservation of fluid implies a radial velocity $v_r = \pm \, r_h^2 / r^2$, where $r_h$ is the radius at which the flow speed exceeds the speed of sound in the fluid. Perturbations $\delta \mathbf{v} = - \nabla \psi$ to the steady flow are governed by Eq.~(\ref{klein-gordon-eqn}) with the effective geometry
\begin{eqnarray}
ds^2 = & g_{\mu \nu} dx^\mu dx^\nu = + d\bar{t}^2 - \left( dr^2 \pm \frac{r_h^2}{r^2} d\bar{t} \right)^2 
\nonumber\\
& - r^2 \left(d \theta^2 + \sin^2 \theta d\phi^2 \right).   \label{geom1}
\end{eqnarray}
The ($+$) sign stands for a source and the ($-$) sign stands for a sink.
We will henceforth choose the ($-$) sign, so that the system is analogous to a black (rather than white) hole. By introducing an alternative time coordinate $dt = d\bar{t} - (r_h^2/r^2) (1-r_h^4/r^4)^{-1} dr$ the metric may be written in a diagonal form,
\begin{equation}
ds^2 = + f(r) dt^2 - f^{-1}(r) dr^2 - r^2 (d\theta^2 + \sin^2 \theta d\phi^2) ,  \label{geom2}
\end{equation}
where $f(r) = 1 - r_h^4 / r^4$. Note that the time parameter $t$ diverges as $r \rightarrow r_h$.

Since the perturbation equation~(\ref{klein-gordon-eqn}) is simply the Klein-Gordon equation on a curved background, we may employ standard field-theory techniques. We allow the field $\psi$ to take complex values, and construct the conserved current
\begin{equation}
J_\mu = i \left(  \psi^\ast \partial_\mu \psi - \psi \partial_\mu \psi^\ast \right)
\end{equation}
satisfying
\begin{equation}
\nabla_\mu \left( g^{\mu \nu} J_\nu \right) = 0 .
\end{equation}
Alternatively, for direct comparison with acoustic experiments we may work directly with the real part of the field $\text{Re}(\psi)$, which corresponds to perturbations in the flow velocity, $\delta \mathbf{v} = - \nabla [ \text{Re}( \psi ) ]$. 

The appropriate choice of observable ($J_\mu$ or $\psi$) depends on the point of view adopted. If we consider acoustic scattering experiments to be a model for black holes irradiated by incoherent electromagnetic radiation, it makes sense to work with $J_\mu$, which behaves like an intensity. If, on the other hand, we are interested in coherent gravitational radiation impinging on a black hole, we should consider the field $\psi$ itself. Interferometers such as LIGO will measure a dimensionless strain $h = \Delta L / L \sim 1/r$ rather than an intensity $I \sim 1/r^2$~\cite{Hughes-2003}. In the remainder of the paper, we will take the former approach, and apply standard techniques from quantum mechanics~\cite{Gottfried-Yan-2004}. 

\section{Geodesics and Orbits\label{sec-orbits}} 
In the short-wavelength limit ($\lambda \ll r_h$), perturbations obey the eikonal approximation $\psi \sim e^{i k^\mu x_\mu}$, where $k^\mu$ is tangent to a null geodesic ($k^\mu k_\mu = 0$) on the background geometry. This naturally motivates a study of the scattering of null geodesics on spherically-symmetric, static backgrounds. For greater insight, let us keep the analysis general and consider the class of geometries described by line elements
\begin{equation}
ds^2 = + f dt^2 - f^{-1} dr^2 - r^2 (d \theta^2 + \sin^2 \theta d\phi^2) , 
\end{equation}
where $ f = 1 - r_h^n / r^n$ and $r_h$ is the horizon radius of the hole. Here, $n=1$ corresponds to the Schwarzschild black hole (a solution of the vacuum Einstein field equations) and $n = 4$ corresponds to the canonical acoustic hole (a solution of fluid-flow equations). 

The paths of scattered geodesics are found from the solutions of the orbital equation 
\begin{equation}
\left(\frac{d u}{d \phi}\right)^2 = \frac{1}{b^2} - u^2 + r_h^n u^{n+2} , \label{orbit-eq-1}
\end{equation}
where $u = 1/r$, and $b$ is the impact parameter. 
Differentiating Eq.~(\ref{orbit-eq-1}) leads to a version of Binet's equation~\cite{dInverno},
\begin{equation}
\frac{d^2u}{d \phi^2} + u = \frac{n+2}{2} r_h^n u^{n+1} .
\label{orbit-eq-2}
\end{equation}
Solutions of Eq.~(\ref{orbit-eq-1}) give the deflection (or scattering) angle as a function of the impact parameter, $\theta(b)$. 
For certain values of $n$, closed-form solutions may be found in terms of elliptic integrals. 
For example, for the canonical acoustic hole ($n=4$), the scattering angle is 
\begin{equation}
\theta(b) = \frac{2 K(k)}{\sqrt{(v_2 - v_1) v_3}}  - \pi, 
\label{scat-ang-sol}
\end{equation}
where
\begin{equation}
k^2 = \frac{v_2 (v_3 - v_1)}{v_3 (v_2 - v_1)}.  \nonumber
\end{equation}
Here, $v_1$, $v_2$ and $v_3$ are the roots of the cubic
\begin{equation}
r_h^4 v^3 - v + 1/b^2 = 0 ,  \label{cubic-roots}
\end{equation}
and $K(k)$ is a complete elliptic integral of the first kind~\cite{Abramowitz-Stegun}. For scattering trajectories, we have $v_1 < 0, v_2 > 0$ and $v_3 > v_2$.

Let us consider approximate solutions of Eq.~(\ref{orbit-eq-1}) in the weak- and strong-field regimes. 
By treating the term on the right-hand side of Eq.~(\ref{orbit-eq-2}) as a small perturbation, it is straightforward to show that, in the weak-field limit, the deflection angle $\theta$ is approximately
\begin{equation}
\theta \approx \alpha_n \frac{r_h^n}{b^n} ,  \label{weak-field-defl}
\end{equation}
where the constant of proportionality is $\alpha_n = \{2, 3\pi/4, 8/3, 15\pi/16, 16/5, 35\pi/32 \}$ for $n = 1 \ldots 6$, respectively. In particular, for the Schwarzschild black hole ($n=1$) we get Einstein's deflection angle $\theta \approx 2 r_h /b$, and for the canonical acoustic hole ($n=4$) we find $\theta \approx 15 \pi r_h^4 / 16 b^4$. 

An unstable circular orbit is present at the critical radius $r_c = \left(\tfrac{n+2}{2}\right)^{1/n} r_h$. This in turn implies a critical impact parameter, $b_c = r_c \left( 1 - r_h^n / r_c^n \right)^{-1/2}$. An incident ray starting with impact parameter $b = b_c$ ends on the unstable orbit at $r = r_c$; the ray circles the hole an infinite number of times. Rays with $b < b_c$ are absorbed by the hole; rays with $b > b_c$ are scattered. The classical absorption cross section is simply the area of a circle of radius $b_c$, that is  $\sigabs = \pi b_c^2$. Numerical values for $r_c$, $b_c$ and $\sigabs$ are listed in Table~\ref{tbl-numerical}.

\begin{table}
\begin{tabular}{lllll}
$n$ \quad \quad & $r_c / r_h$ \quad \quad & $b_c / r_h$ \quad \quad& $\sigabs / \pi r_h^2$ \quad \quad \\ 
$1$ & $1.5000$ & $2.5981$ & $6.7500$ \\ 
$2$ & $1.4142$ & $2.0000$ & $4.0000$ \\ 
$3$ & $1.3572$ & $1.7522$ & $3.0700$ \\ 
$4$ & $1.3161$ & $1.6119$ & $2.5981$ \\ 
$5$ & $1.2847$ & $1.5201$ & $2.3108$ \\ 
$6$ & $1.2599$ & $1.4548$ & $2.1165$ \\ 
\end{tabular}
\caption{Numerical values of the unstable orbit radius ($r_c$), the critical impact parameter ($b_c$) and the absorption cross section ($\sigabs$) for $n = 1\ldots 6$, where $n=1$ corresponds to the Schwarzschild black hole, and $n = 4$ corresponds to the canonical acoustic hole.}
\label{tbl-numerical}
\end{table}

Geodesics passing very close to the unstable circular orbit may be scattered through large angles. For $b \gtrsim b_c$, the deflection angle is approximately
\begin{equation}
\theta(b) \approx - \frac{1}{\sqrt{n}} \ln \left( \frac{b - b_c}{\beta_n b_c} \right) .  \label{strong-field-defl}
\end{equation}
Here $\beta_n$ is a numerical coefficient which may be calculated on a case-by-case basis. 
Many years ago, Darwin~\cite{Darwin-1959} showed that $\beta_1 \approx 0.6702$ for the Schwarzschild black hole. By considering the asymptotic form of Eq.~(\ref{scat-ang-sol}) as $k \rightarrow 1$, it is straightforward to show that
$\beta_4 = 108 e^{-2 \pi} \approx 0.2017$ for $n=4$ (the canonical acoustic hole). The calculation is outlined in Appendix~\ref{appendix-spiral}.

It is well-known that an unstable orbit gives rise to a \emph{glory} in the backward direction. A glory is a bright spot or ring in the backward scattering direction. The size and intensity of the spot or ring depends on the wavelength. Via path-integral arguments it was shown that~\cite{Matzner-1985}, close to $\theta = \pi$, the scattering cross section is approximately
\begin{equation}
\frac{d \sigma}{d \Omega} \approx 2 \pi \omega b_g^2 \left| \frac{d b}{d \theta} \right|_{\theta=\pi} \, \left[ J_{2s} (b_g \omega \sin \theta) \right]^2\,. \label{glory-1}
\end{equation}
Here $b_g$ is the impact parameter for which the deflection angle is $\pi$, $J_{2s}(x)$ is a Bessel function, and $s$ is the spin of the scattered field (in this case $s = 0$). 

Combining results~(\ref{strong-field-defl}) and~(\ref{glory-1}) leads to the glory approximation
\begin{equation}
{r_h}^{-2} \frac{d \sigma}{d \Omega} \approx 0.019835 \, \omega r_h  \left[ J_{0} ( 1.61246 \, \omega r_h \sin \theta ) \right]^2  \label{glory-approx}
\end{equation}
for the canonical acoustic hole~\cite{footnote1}, 
which may be compared with the approximation
\begin{equation}
{r_h}^{-2}  \frac{d \sigma}{d \Omega} \approx 3.3772 \, \omega r_h \left[ J_{0} ( 2.67325 \, \omega r_h \sin \theta) \right]^2 \label{schw-glory-approx}
\end{equation}
for the Schwarzschild black hole. In other words, the glory peak for the canonical acoustic hole is similar in nature to the Schwarzschild black hole, but is approximately $170$ times weaker in magnitude~\cite{footnote2}. For a given frequency, the angular width of the glory peak of the Schwarzschild black hole is narrower than for the canonical acoustic hole, by a factor of approximately $1.66$.  In the next section, we test the validity of approximation~(\ref{glory-approx}) by computing the exact cross section via partial wave series. 

\section{Partial Wave Method\label{sec-partialwave}}
In this section we use the partial wave method to determine the differential scattering cross section $d\sig / d\Omega$ at intermediate wavelengths ($\lambda \sim r_h$). 

By substituting the metric~(\ref{geom2}) and the separation ansatz
\begin{equation}
\psi = \frac{u(r)}{r} Y_{lm}(\theta, \phi) e^{-i \omega t} 
\end{equation}
into Eq.~(\ref{klein-gordon-eqn}) we obtain a radial equation
\begin{equation}
f \frac{d}{d r} \left(  f \frac{d u}{d r} \right) + \left[ \omega^2 - V_l(r) \right] u = 0, 
\label{radial-eq}
\end{equation}
where
\begin{equation}
V_l(r) = f \left( \frac{f^\prime}{r} + \frac{l(l+1)}{r^2} \right) , \label{Veff}
\end{equation}
and $f^\prime = df/dr$.  
The perturbation must be purely ingoing at the horizon $r = r_h$. Hence we impose the boundary condition
\begin{equation}
u(r) = \exp( - i \omega x) \quad  \quad \quad \text{as} \quad r \rightarrow r_h , \label{bc1}
\end{equation}
where
\begin{equation}
\frac{d x}{dr} = f^{-1} , 
\nonumber
\end{equation}
so that
\begin{equation}
x = r + \frac{r_h}{4} \ln \left| \frac{r - r_h}{r + r_h} \right| - \frac{1}{2} r_h \tan^{-1}\left(\frac{r}{r_h}\right) + \frac{r_h \pi}{4} ,
\nonumber
\end{equation}
is the Wheeler-type coordinate. The constant of integration has been chosen so that $x \approx r$ as $r \rightarrow \infty$. 
It is straightforward to check that in the original coordinate system~(\ref{geom1}) the ingoing solution $e^{-i\omega x}$ is well-defined at $r = r_h$, whereas the outgoing solution $e^{+i\omega x}$ is divergent. Towards spatial infinity, the asymptotic form of the solution is
\begin{equation}
u_l(r) \sim \omega x \left[   \Aout  \, i^{l+1} h_l^{(1)}(\omega x)  +  \Ain \, (-i)^{(l+1)} h_l^{(1) \ast}(\omega x)   \right], \label{bc2} 
\end{equation}
where $h_{l}^{(1)} (x)$ are the spherical Bessel functions of the third kind~\cite{Abramowitz-Stegun}, and $\Ain$ and $\Aout$ are complex constants. 

The phase shifts $\delta_l$ are defined by
\begin{equation}
e^{2i\delta_l} = (-1)^{l+1} \Aout / \Ain   \label{phase-shift-def}
\end{equation}
and the scattering amplitude is
\begin{equation}
f(\theta) = \frac{1}{2 i \omega} \sum_{l=0}^\infty (2l+1) (e^{2 i \delta_l} - 1) P_l(\cos \theta) .  \label{f-scat}
\end{equation}
From the amplitude, the differential scattering cross section follows immediately,
\begin{equation}
\frac{d \sigma}{d \Omega} = \left| f(\theta) \right|^2 . \label{diff-scat-csec}
\end{equation}

Finally, we can define the scattering (elastic), absorption (inelastic) and total (combined) cross sections~\cite{Gottfried-Yan-2004}, 
\begin{eqnarray}
\sigscat 
&\equiv&  \int  \frac{d\sig}{d\Omega}  d \Omega \nonumber \\
&=& 
\frac{\pi}{\omega^2} \sum_{l=0}^\infty (2l+1) \left| e^{2i\delta_l} - 1 \right|^2\, ;  \label{sig-el} \\
\sigabs &=& \frac{\pi}{\omega^2} \sum_{l=0}^\infty (2l+1) \left( 1 - \left| e^{2i\delta_l} \right|^2 \right)\, ; \label{sig-inel}  \\
\sigtot 
&\equiv&  \sigscat+ \sigabs \nonumber \\
&=& \frac{2 \pi}{\omega^2} \sum_{l=0}^\infty (2l+1) \left( 1 - \text{Re}\left( e^{2i\delta_l} \right) \right)\, . \label{sig-tot}
\end{eqnarray}

To compute the cross sections~(\ref{diff-scat-csec}--\ref{sig-tot}) we must solve Eq.~(\ref{radial-eq}) subject to the boundary conditions~(\ref{bc1}) and~(\ref{bc2}), to obtain numerical values for the phase shifts via Eq.~(\ref{phase-shift-def}). The numerical method employed is described in section~\ref{sec-numerical}. The task is made substantially easier if we utilise an approximation for the phase in the large-$l$ limit. 
We find that, for $l \gg l_c \sim \omega b_c$, the phase shift for the scattering from the canonical acoustic hole is approximately
\begin{equation}
\delta_l \approx \frac{5 \pi (\omega r_h)^4}{32 (l+1/2)^3} .  \label{phaseshift-approx}
\end{equation}
There are (at least) two ways to reach this result. 

The first way to obtain Eq.~(\ref{phaseshift-approx}) is through Ford and Wheeler's~\cite{Ford-1959} semiclassical description of scattering,  whereby an impact parameter is associated with each partial wave, 
\begin{equation}
b = \frac{  l + 1/2  }{  \omega } .
\end{equation}
Much physical information can be extracted from the deflection function $\Theta(l)$, defined by
\begin{equation}
\Theta(l) = \frac{d}{dl}\left[  \text{Re} (2 \delta_l)   \right] . \label{defl-fn}
\end{equation}
Here, $l$ is allowed to assume continuous real values. Equating the deflection function $\Theta$~(\ref{defl-fn}) with the weak-field deflection angle $\theta$~(\ref{weak-field-defl}) yields
\begin{equation}
\frac{d}{dl} \left[ \text{Re} (2 \delta_l) \right]  = \frac{15 \pi}{16} \frac{r_h^4}{b^4} = \frac{15 \pi}{16} \frac{(\omega r_h)^4}{(l+1/2)^4}\,,
\end{equation}
from which Eq.~(\ref{phaseshift-approx}) follows immediately (up to a sign).

The second way to obtain Eq.~(\ref{phaseshift-approx}) is through the Born approximation~\cite{Gottfried-Yan-2004}. First, we make the substitution $u(r) = f^{-1/2} X(r) $ in Eq.~(\ref{radial-eq}), so that 
\begin{equation}
\frac{d^2 X}{d r^2} + f^{-2} \left[ \omega^2 - V_l(r) - \frac{f}{2} \frac{d^2f}{dr^2} + \frac{1}{4} \left( \frac{df}{dr} \right)^2 \right] X = 0.
\end{equation}
Next, we approximate the wave equation as a power series in $1/r$,
\begin{equation}
\frac{d^2 X}{d r^2} + \left[ \omega^2 - \frac{l(l+1)}{r^2} + U(r)   \right] X = 0 ,
\end{equation}
where
\begin{equation}
U(r) = \frac{2 r_h^4 \omega^2}{r^4} - \frac{(l(l+1) - 6)r_h^4}{r^6} + \frac{3 r_h^8 \omega^2}{r^8} + \mathcal{O}\left( \frac{1}{r^{10}} \right) . 
\end{equation}
Now we apply the Born approximation formula~\cite{Morse-Feshbach},
\begin{equation}
\delta_l \approx - \omega \int_0^\infty r^2 j_l^2( \omega r) U(r) dr ,
\end{equation}
where $j_l(x)$ are the spherical Bessel functions of the first kind~\cite{Abramowitz-Stegun}.  The first two terms in $U(r)$ give a contribution proportional to $(\omega r_h)^4$. Applying the identities
\begin{align}
&\int_0^\infty x^{-2} j_l^2(x) dx  = \frac{ \pi }{(2l-1)(2l+1)(2l+3)}  \\
&\int_0^\infty x^{-4} j_l^2(x) dx =  \nonumber \\
&\frac{ 3\pi }{ (2l-3)(2l-1)(2l+1)(2l+3)(2l+5) } 
\end{align}
leads to Eq.~(\ref{phaseshift-approx}) to lowest order in $l+1/2$. 

\section{Numerical Method\label{sec-numerical}}
To compute numerical phase shifts, we first solve the radial equation~(\ref{radial-eq}) subject to Eq.~(\ref{bc1}) and compute ingoing and outgoing coefficients ($\Ain$ and $\Aout$, respectively) by matching on to Eq.~(\ref{bc2}). The phase shifts follow immediately from Eq.~(\ref{phase-shift-def}). 

Our numerical method is straightforward. We start close to the horizon, at $r = r_h + \eta \, r_h$ (where typically $\eta \sim 0.001$) with a series expansion
\begin{equation}
\psi = e^{-i\omega x } \sum_{k = 0}^{k_m} a_k \eta^k .
\end{equation}
The coefficients $a_k$ can be found with a symbolic algebra package such as Maple or Mathematica. We evaluated the series at 6th order ($k_m=6$). Next, we numerically integrate from near the horizon into the asymptotically flat region, $r = r_m \sim 150 r_h$. We matched the numerical solution onto Eq.~(\ref{bc1}). We checked the accuracy of the resulting phase shifts by varying $\eta$ and $r_m$.

\begin{figure}[htb!]
\centering
\includegraphics[scale=0.65]{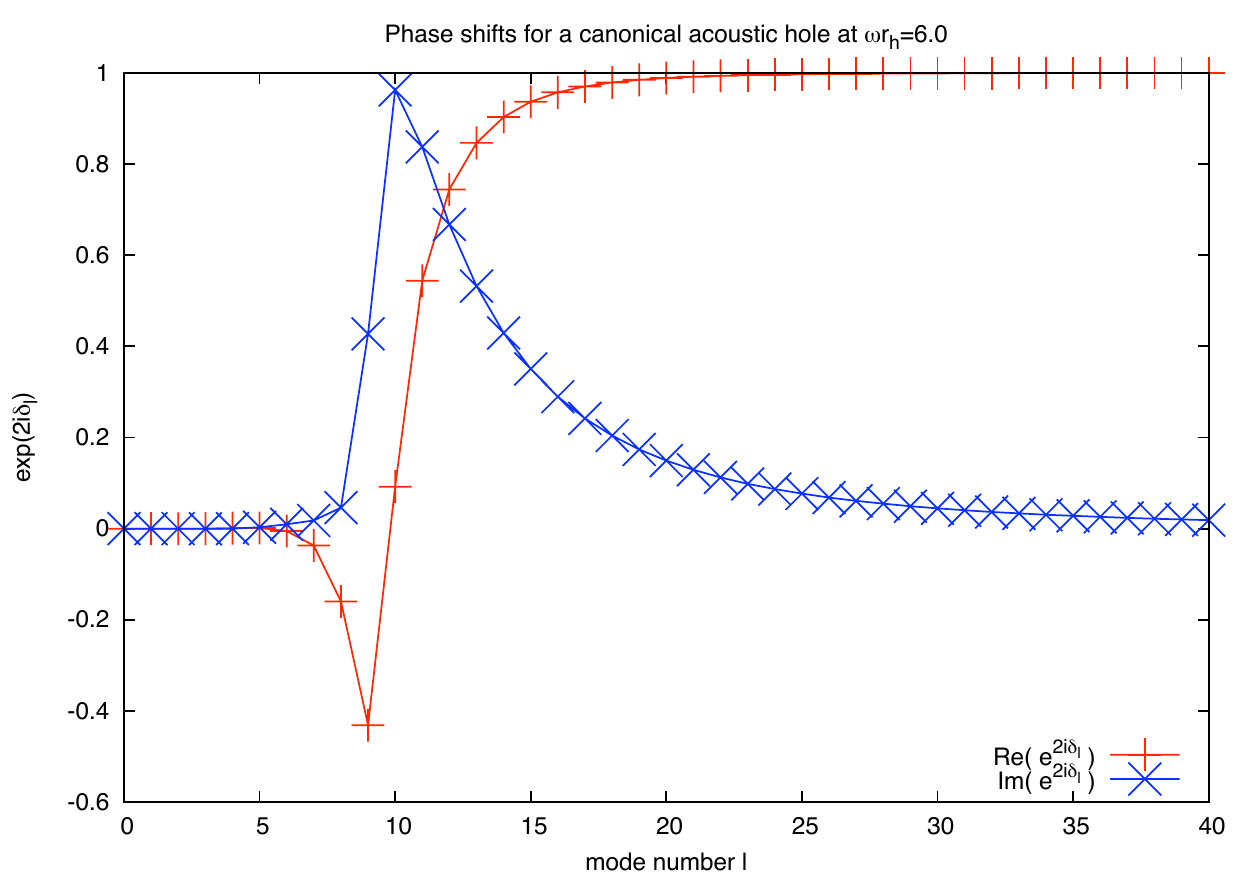}
 \caption{\emph{Phase shifts for a canonical acoustic hole at $\omega r_h = 6.0$}. Below $l_c \sim \omega b_c \approx 1.612 \omega r_h \approx 9.7$, absorption dominates. In the large-$l$ limit, $e^{2i \delta_l} \sim 1 + 5 i \pi (\omega r_h)^4  / 16 (l+1/2)^{3}$.}
 \label{fig-phases-w6}
\end{figure}

Fig.~\ref{fig-phases-w6} shows typical phase shifts as a function of angular mode number $l$. The real  and imaginary components of $e^{2i \delta_l}$ are plotted. It is clear that partial waves of low order  ($l \ll l_c = \omega b_c$) are almost entirely absorbed. Partial waves of intermediate order ($l \sim l_c$) are partially scattered and absorbed. In this regime the phase shift varies rapidly with $l$. For partial waves of large order ($l \gg l_c$) the phase shift $\delta_l$ is purely real, and conforms to approximation~(\ref{phaseshift-approx}). 

In practice, we must truncate the partial wave sum in Eq.~(\ref{f-scat}) at some finite upper limit $l = \lmax$. If done naively, this introduces a truncation error. To avoid this problem, we employed the asymptotic phase shifts~(\ref{phaseshift-approx}) to minimize the loss of precision. The amplitude may be split into two parts,
\begin{eqnarray}
f(\theta) &=& f_{num}^{(\lmax)}(\theta) + f_{rem}^{(\lmax)}(\theta) \nonumber \\
&=& \frac{1}{2i\omega}\sum_{l=0}^{\lmax} (2l+1) (e^{2i\delta_l} - 1) P_l(\cos \theta)  \nonumber  \\
&+&   \left(  \frac{5\pi \omega^3 r_h^4}{16} \right) \sum_{l=\lmax + 1}^\infty \frac{P_l(\cos\theta)}{(l+1/2)^2} .
\end{eqnarray}
The remainder term $f_{rem}$ may either be approximated numerically (by choosing some extremely large cutoff) or, in certain cases, computed analytically. For example, to compute the sum in the forward direction $\theta = 0$, for which $P_l(1) = 1$, we may use the result
\begin{equation}
\sum_{l = \lmax+1}^\infty (l+1/2)^{-2} = \Psi( 1, \lmax + 3/2 ) ,
\end{equation}
where $\Psi$ is the polygamma function~\cite{Abramowitz-Stegun}. 
To compute the remainder sum in the backward direction $\theta = \pi$, for which $P_l(-1) = (-1)^l$ we may use the result
\begin{equation}
\sum_{l = 0}^\infty (-1)^l (l+1/2)^{-2}  =  4 \, \mathcal{C} \,,
\end{equation}
where $\mathcal{C} \approx 0.91596559$ is Catalan's constant~\cite{Gradshteyn-Ryzhik}.

\section{Results\label{sec-results}}
In this section, we present our numerical results and compare with theoretical approximations and expectations.

\begin{figure}[htb!]
\centering
\includegraphics[scale=0.65]{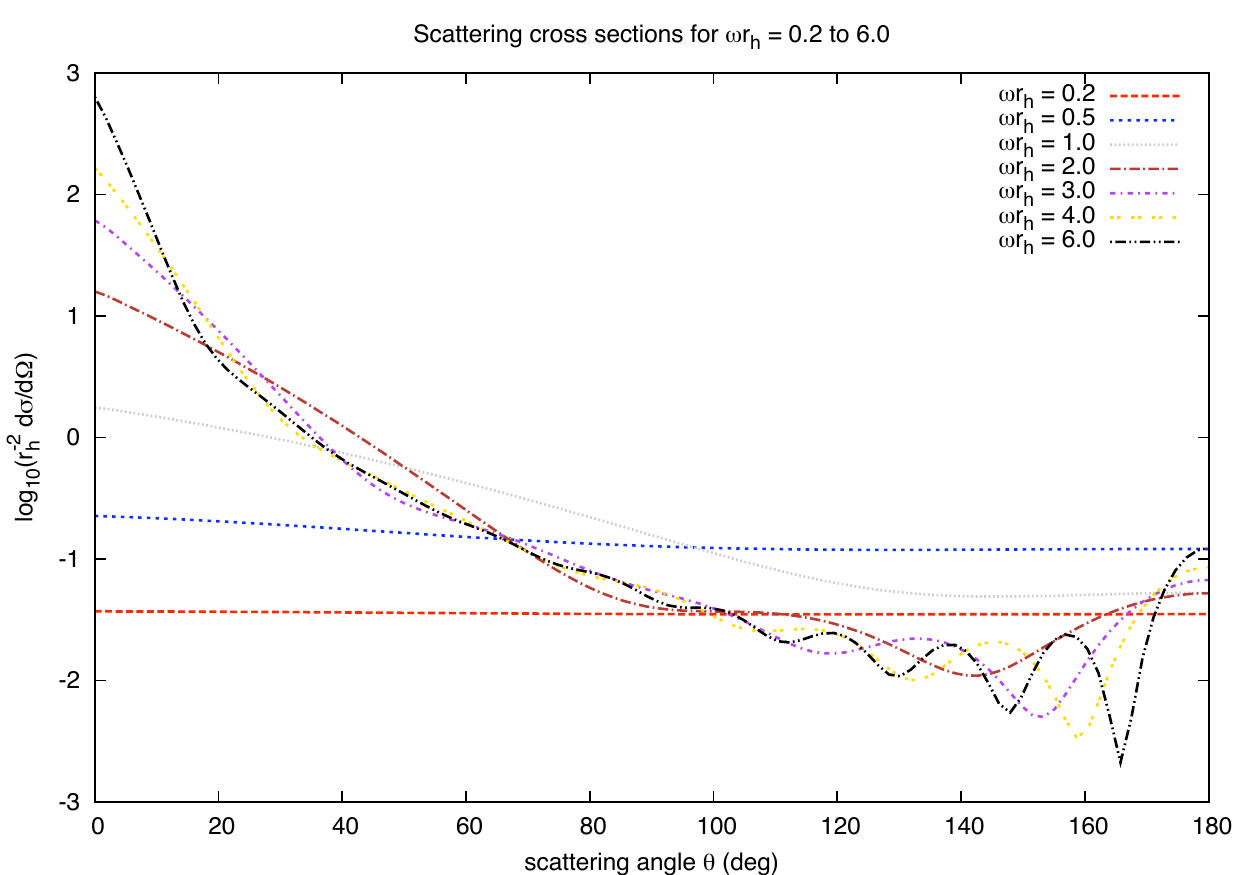}
 \caption{\emph{Scattering pattern of a canonical acoustic hole for a range of incident wavelengths}.  The plot shows the (logarithm of the) differential scattering cross section  $d\sigma / d\Omega$ as a function of scattering angle $\theta$ for a range of frequencies $\omega r_h$. }
 \label{fig-om0to6}
\end{figure}

Fig.~\ref{fig-om0to6} shows the scattering cross section as a function of angle, for a range of couplings (for $\omega r_h = 0.2 \,\, \text{to}\,\, 6.0$). At small couplings ($\omega r_h \ll 1$), the scattered flux is isotropic, and the $l=0$ mode dominates. At higher couplings, flux is preferentially scattered in the forward direction, and a more complicated pattern arises. In particular, for $\omega r_h \gtrsim 1$ we see an oscillatory pattern with an angular width inversely proportional to $\omega r_h$. Similar patterns are observed for black hole scattering~\cite{Futterman-1988, Glampedakis-Andersson-2001, Dolan-2006, Dolan-2008}. An explanation for the physical origin of the oscillations can be found in~\cite{Matzner-1985, Anninos-1992}. The oscillations arise from the interference of rays that pass in opposite senses around the hole (i.e. interference between rays scattered through $\theta$, $2\pi - \theta$, $2 \pi + \theta$, $\ldots$ etc.~\cite{Laven-2005}).
 
\begin{figure}[htb!]
\centering
\includegraphics[scale=0.65]{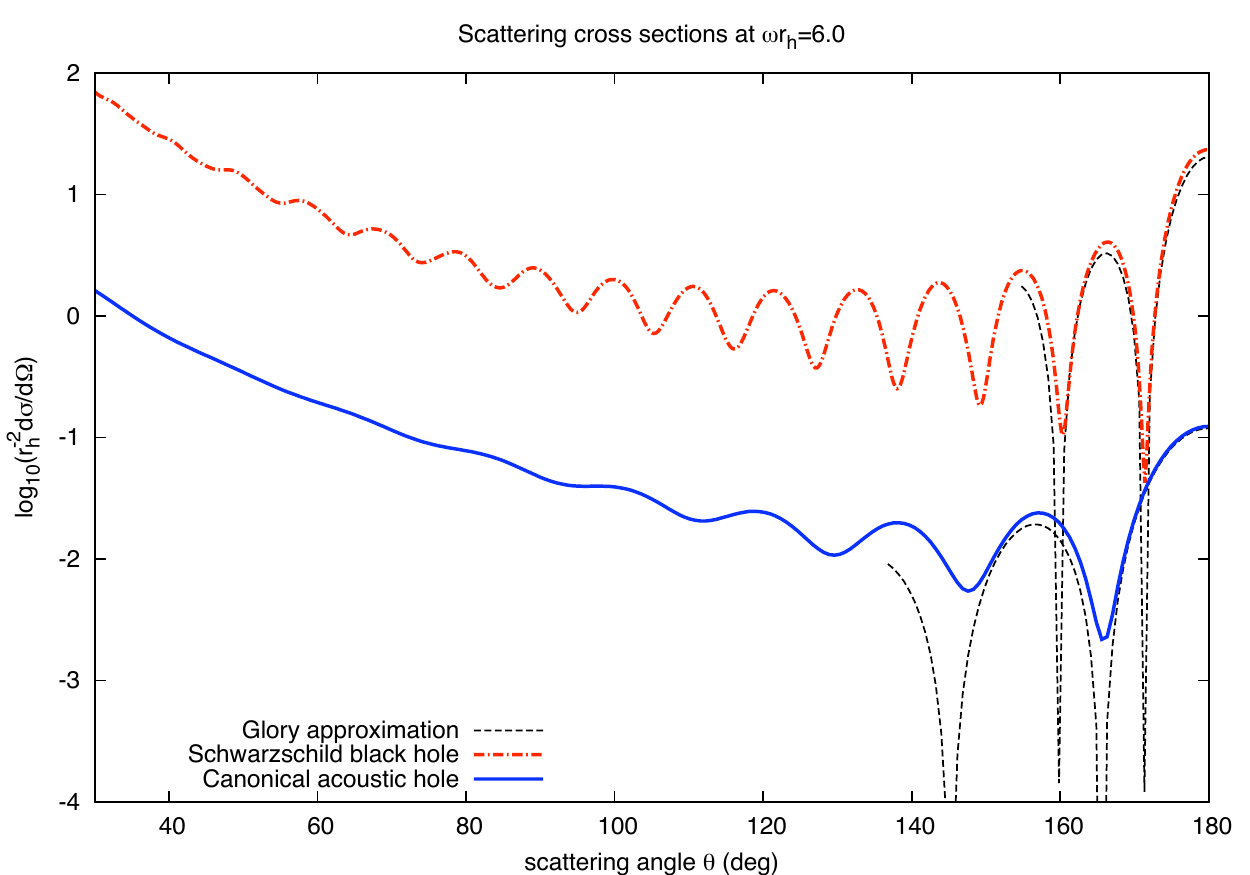}
 \caption{\emph{Scattering cross section of Schwarzschild black hole and canonical acoustic hole at $\omega r_h = 6.0$}. The numerically-determined scattering cross sections are compared with the glory approximations (Eqs.~(\ref{glory-approx}) and~(\ref{schw-glory-approx})). Note the logarithmic scale on the vertical axis.}
 \label{fig-glory-w6}
\end{figure}

Fig.~\ref{fig-glory-w6} shows an example of the interference effects at $\omega r_h = 6.0$. The plot compares the scattering cross section for the Schwarzschild black hole with the canonical acoustic hole. The Schwarzschild black hole scatters much more flux, due to the long-range nature of the interaction ($\phi \sim 1/r$ versus $\phi \sim 1/r^4$). In addition, the oscillations from the Schwarzschild black hole are narrower than from the canonical acoustic hole, by a factor of approximately $1.66$. The glory approximations~(\ref{glory-approx}) and~(\ref{schw-glory-approx}) are shown as broken lines. 

\begin{figure}[htb!]
\centering
\includegraphics[scale=0.65]{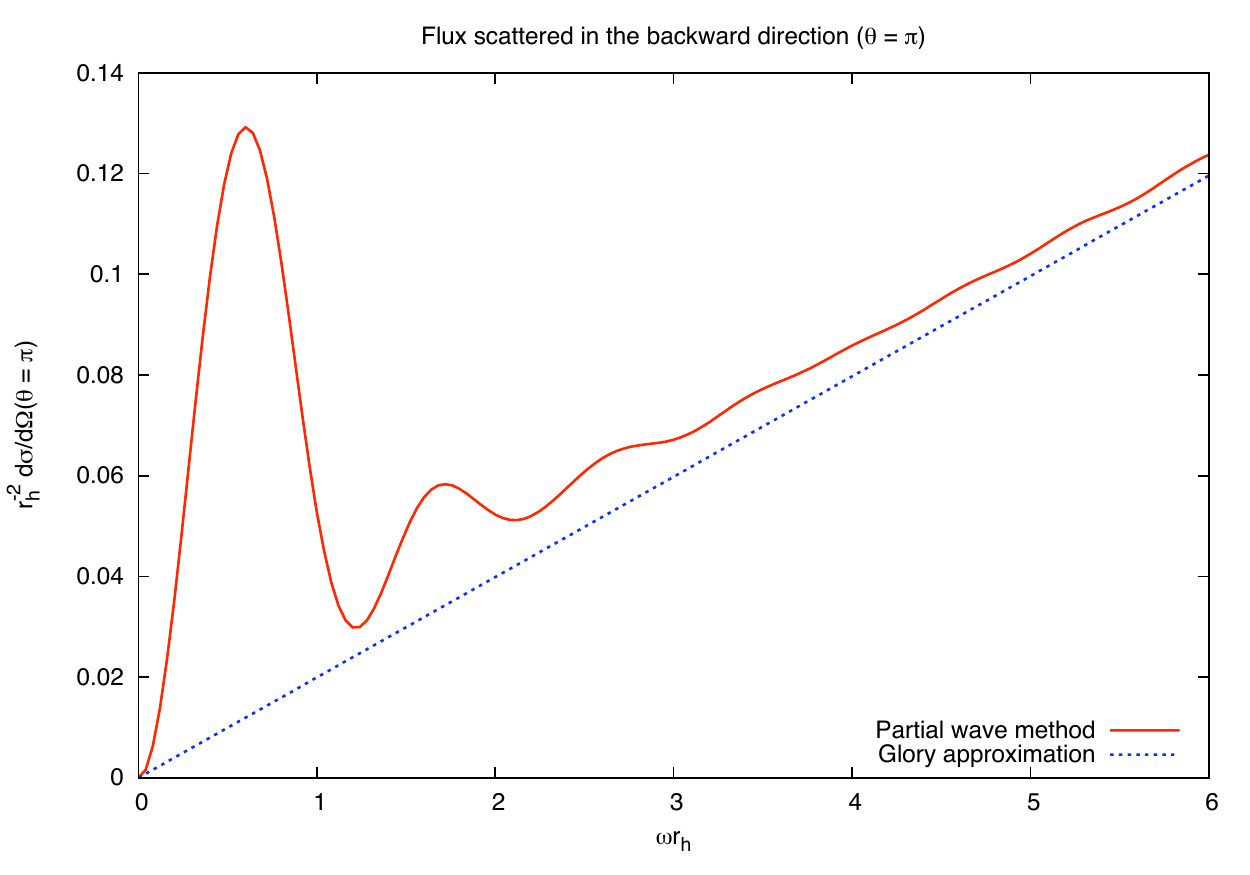}
 \caption{\emph{Flux scattered in the backward direction}. The plot shows $d\sigma / d\Omega$ at $\theta = 180^\circ$ as a function of the coupling $\omega r_h$. }
 \label{fig-180}
\end{figure}

Fig.~\ref{fig-180} shows the differential scattering cross section in the backward direction, $\theta = 180^\circ$, as a function of frequency. At low frequencies, the back-scattered flux is predominantly due to the $l=0$ mode. At high frequencies, the glory approximation predicts that the flux increases linearly with $\omega r_h$. The glory approximation is shown as a straight line, and provides a good fit to the numerical data.

We note that for the canonical acoustic hole, the scattering cross section in the forward direction ($\theta = 0^\circ$) is well-defined and finite (see Fig.~\ref{fig-om0to6}). This contrasts with the Schwarzschild black hole, for which $d\sig / d\Omega \sim \theta^{-4}$ as $\theta \rightarrow 0$. It also contrasts with the asymptotic behavior of the `classical' scattering cross section, which may be defined with reference to parallel geodesics approaching from infinity,
\begin{equation}
\left. \frac{d \sigma}{d \Omega} \right|_{\text{cl.}} \approx  \frac{b}{\sin \theta} \left| \frac{d b}{d \theta} \right| \, .
\end{equation} 
Substituting in the weak-field deflection angle~(\ref{weak-field-defl}) leads to $d \sig / d \Omega|_{\text{cl.}} \sim \theta^{-5/2} $.

\begin{figure}[htb!]
\centering
\includegraphics[scale=0.65]{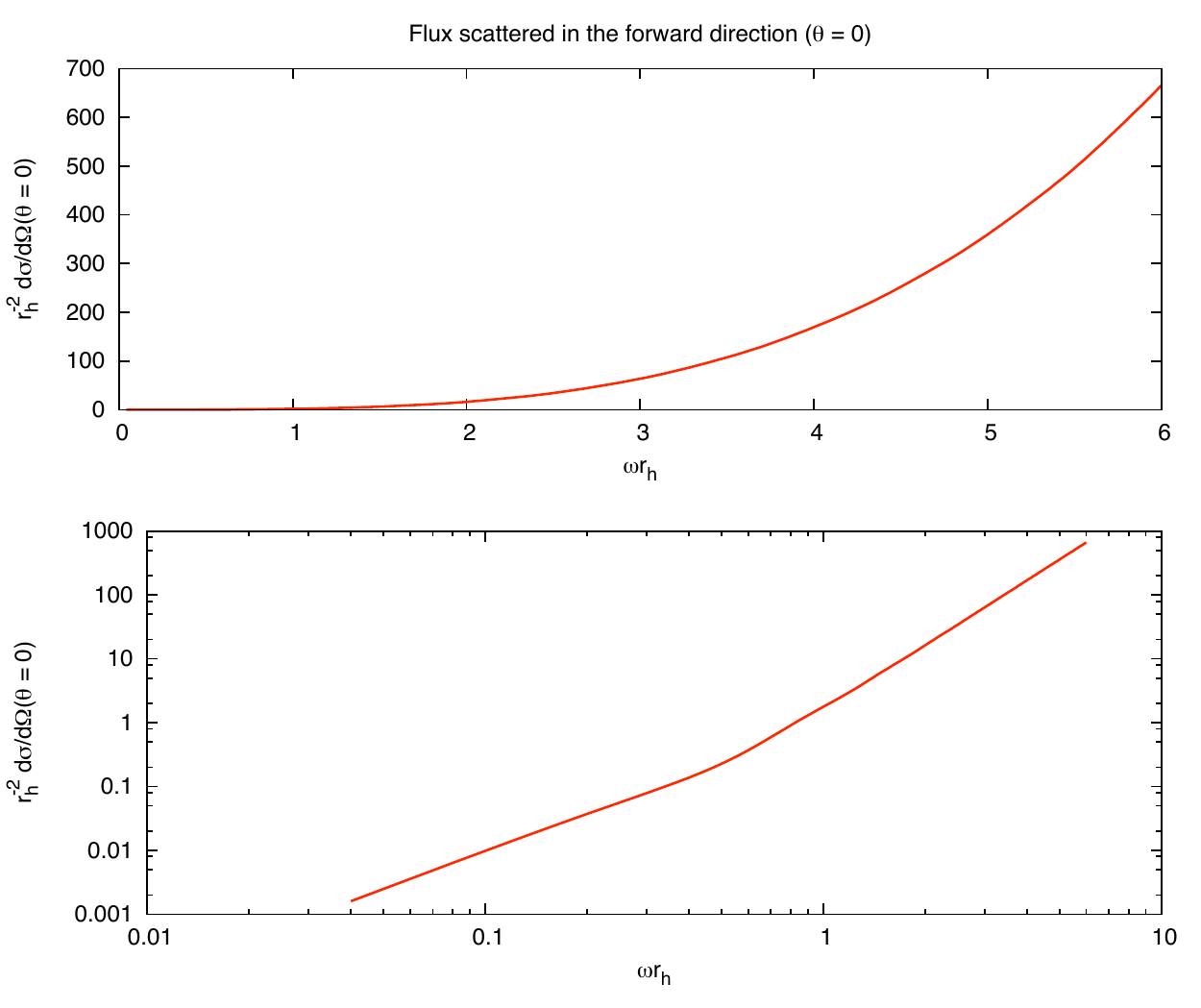}
 \caption{\emph{Flux scattered in the forward direction}. The plots show $d\sigma / d\Omega$ at $\theta = 0^\circ$ as a function of the coupling $\omega r_h$. The lower plot shows the same data as the upper plot, but with a log-log scale. The data suggests that the on-axis cross section diverges as $(\omega r_h)^\gam$ with an exponent $3 < \gam < 4$, as $\omega r_h \rightarrow \infty$.}
 \label{fig-0}
\end{figure}

The scattering cross section in the forward direction ($\theta = 0^\circ$) as a function of frequency is shown in Fig.~\ref{fig-0}. It increases rapidly and monotonically with $\omega r_h$. However, we should not forget that Eq.~(\ref{klein-gordon-eqn}) is only valid for \emph{small} perturbations to the background flow. Hence it is likely the linear approximation will break down on axis. We would expect non-linear (second-order) effects to become important close to $\theta = 0^\circ$. 

\begin{figure}[htb!]
\centering
\includegraphics[scale=0.65]{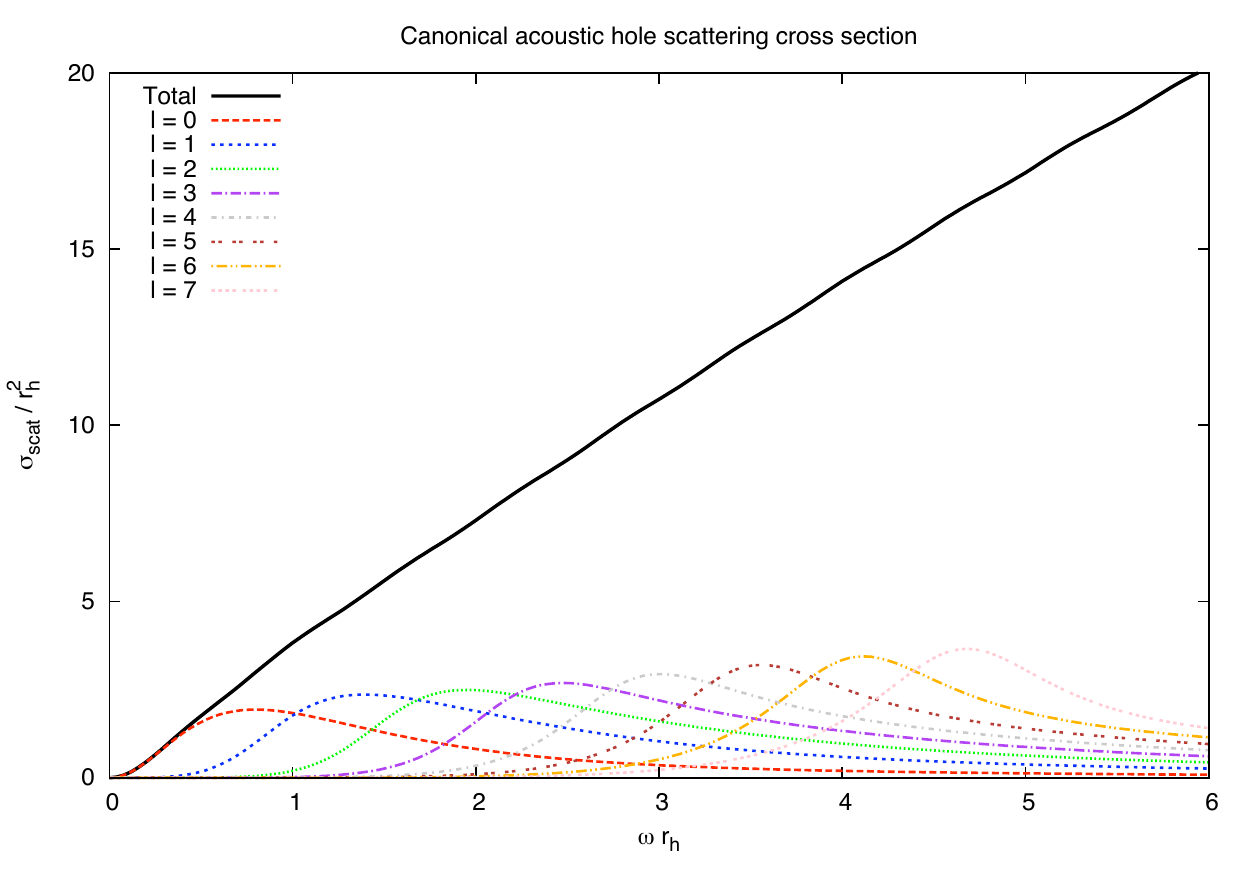}
 \caption{\emph{ Scattering cross section for $0 < \omega r_h < 6$.} The scattering cross section for the canonical acoustic hole, given by Eq.~(\ref{sig-el}), is plotted as a function of the coupling $\omega r_h$. The contributions from the $l = 0 \ldots 7$ partial waves are shown.}
 \label{fig-scat}
\end{figure}

\begin{figure}[htb!]
\centering
\includegraphics[scale=0.65]{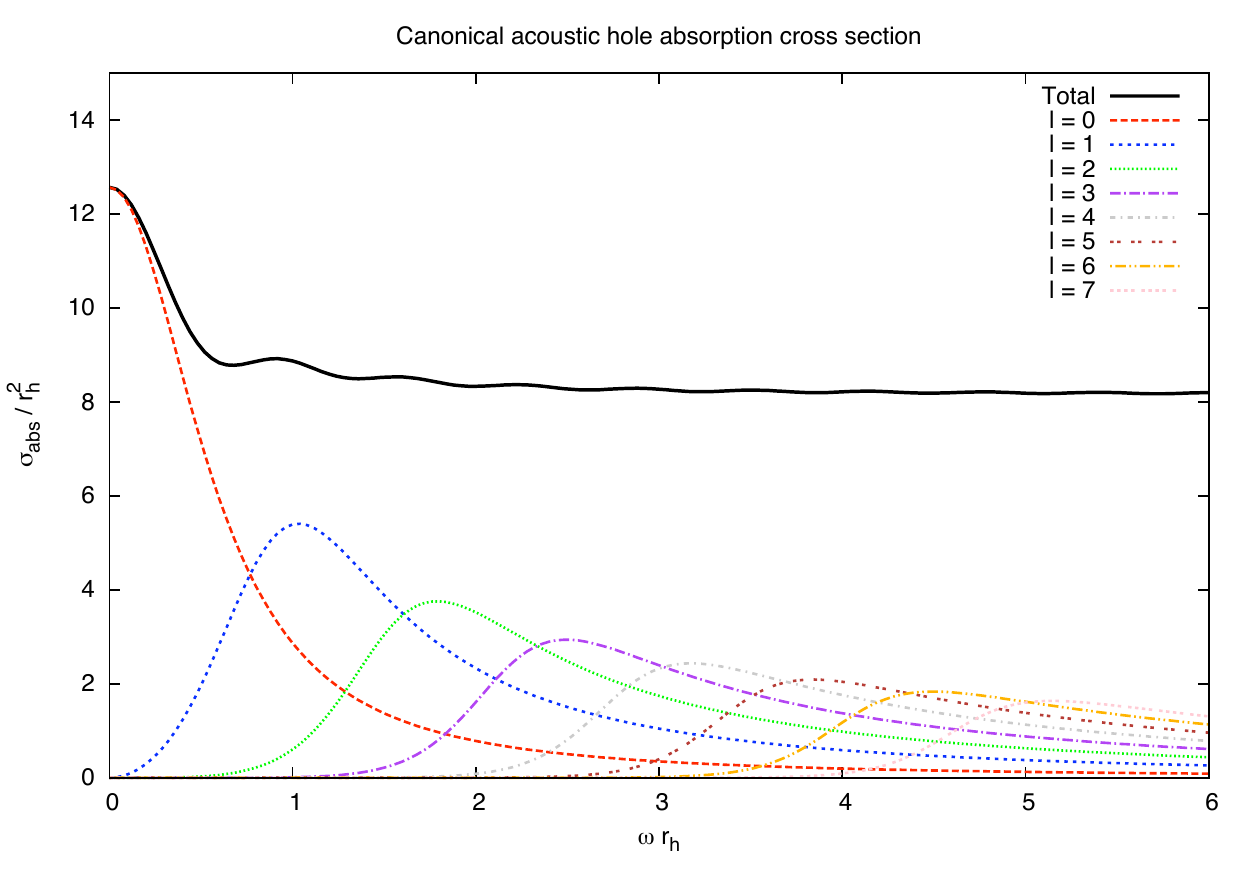}
 \caption{\emph{ Absorption cross section for $0 < \omega r_h < 6$.} The canonical acoustic hole absorption cross section given by Eq.~(\ref{sig-inel}) is plotted as a function of the coupling $\omega r_h$. The contributions from $l = 0$ up to $7$ are also shown.}
 \label{fig-abs}
\end{figure}

The overall scattering and absorption cross sections (defined in Eqs.~(\ref{sig-el}) and~(\ref{sig-inel})) are shown in Fig.~\ref{fig-scat} and in Fig.~\ref{fig-abs}, respectively, for $0 < \omega r_h < 6$. As has been observed previously~\cite{Crispino-Oliveira-Matsas-2007}, the absorption cross section tends to the horizon area $\sigabs \approx 4 \pi r_h^2 $ at low frequencies~\cite{Das-Gibbons-Mathur-1997}, and to the geometric-optics value $\sigabs \approx \pi b_c^2$ at high frequencies. On the other hand, the scattering cross section increases with $\omega r_h$, in an approximately linear fashion.

\section{Final Remarks\label{sec-finalremarks}}
In this paper we have presented an investigation of the scattering of
sound waves by canonical acoustic holes. We have applied analytic and
numerical methods to determine the differential scattering cross
section for a range of wavelength-to-horizon ratios of a monochromatic
planar perturbation impinging on the canonical acoustic hole. We obtained
high-precision results for the scattering cross section with the
partial wave method, considering all angular momentum contributions.

The diffraction pattern created by a canonical acoustic hole is
distinctive (Fig.~\ref{fig-om0to6}). It shares some of the features of
Schwarzschild black hole diffraction \cite{Dolan-2008}. There are
oscillations in the scattering amplitude (Fig.~\ref{fig-glory-w6}) due
to the interference of wavefronts passing in opposite senses around
the hole (so-called \emph{orbiting}~\cite{Anninos-1992}). The
structure of these oscillations is intimately linked to the properties
of the unstable null orbit which lies close to the horizon.

Experimentally, such interference effects would be manifest as angular
segments of alternately weak and strong intensity, whose angular width
is proportional to wavelength. We expect a particularly strong segment
of positive interference in the backward direction: the so-called
\emph{glory}. In this paper we computed the glory via semi-classical
approximations (Eq.~(\ref{glory-approx})) and partial wave methods
(Fig.~\ref{fig-glory-w6}), and found the two approaches to be
consistent (Fig.~\ref{fig-180}).

Wave scattering by canonical acoustic holes is different to wave
scattering by astrophysical black holes in three key respects. Firstly, at long wavelengths ($\omega r_h \ll 1$), scattering by the canonical acoustic hole is isotropic (Fig.~\ref{fig-om0to6}) whereas scattering by the Schwarzschild black hole is Coulombian, i.e.~$d\sigma / d\Omega \approx r_h^2 / 4 \sin^4 (\theta/2)$ \cite{Futterman-1988}.
Secondly, for equivalent couplings $\omega r_h$~\cite{footnote2}, the intensity
of the flux scattered by a canonical acoustic hole is much weaker than the
intensity scattered by a Schwarzschild black hole, by approximately two
orders of magnitude at large angles 
(Fig.~\ref{fig-glory-w6}). Thirdly, the scattering amplitude for the
canonical acoustic hole is finite in the forward direction
(Fig.~\ref{fig-om0to6}), whereas the amplitude for the Schwarzschild
black hole diverges as $\theta \rightarrow 0$
\cite{Futterman-1988}. These differences are explained by the behavior
of the effective potential (\ref{Veff}) far from the hole. The
potential of the canonical acoustic hole decays more rapidly ($\sim 1/r^4$) than
the `Newtonian' potential ($\sim 1/r$). 

Given the exciting prospects for gravitational wave astronomy
\cite{LIGO-2008}, time-domain simulations of wave scattering
by acoustic holes would be of interest. Even better would be
 an experimental realization of wave scattering by the canonical
acoustic hole, or indeed, by any of the alternative black hole
analogues \cite{Barcelo-Liberati-Visser-2005, 
Unruh-1981, Visser-1998, Volovik-2003, Garay-2002, Schutzhold-Unruh-2004, Corley-Jacobson-1999, Unruh-Schutzhold-2007, Novello-Visser-Volovik}. We propose this as a challenge for experimental groups worldwide.

\begin{acknowledgments}
  The authors would like to thank Conselho Nacional de Desenvolvimento
  Cient\'\i fico e Tecnol\'ogico (CNPq) for partial financial support
  and G.~Matsas for helpful discussions. S. D. thanks M. Casals,
  V.~Cardoso, and P.~Watts for discussions, and the Universidade
  Federal do Par\'a (UFPA) in Bel\'em for kind
  hospitality. S. D. acknowledges financial support from the Irish
  Research Council for Science, Engineering and Technology (IRCSET).
  E. O. and L. C. would like to acknowledge also partial financial
  support from Coordena\c{c}\~ao de Aperfei\c{c}oamento de Pessoal de
  N\'\i vel Superior (CAPES).

\end{acknowledgments}

\appendix

\section{Logarithmic Scattering Approximation\label{appendix-spiral}}
Here the constant $\beta_n$ in the spiral scattering approximation~(\ref{strong-field-defl}) is computed for the special case $n = 4$ (the canonical acoustic hole). We start with the exact solution~(\ref{scat-ang-sol}). The roots $v_1, v_2, v_3$ are defined by the cubic~(\ref{cubic-roots}). 
The critical orbit occurs when $r_h^2 v_1 = -2 / \sqrt{3}$, and $r_h^2 v_2 = r_h^2 v_3 = 1/\sqrt{3}$, and $b = b_c = (3^{3/4} / 2^{1/2}) r_h$. For orbits close to the critical orbit, the roots are
\begin{equation}
\begin{array}{lllll}
r_h^2 v_1 & \approx & -2 / 3^{1/2} & & + \quad \mathcal{O} (\delta^2) \\
r_h^2 v_2 & \approx & \;\;\, 1/3^{1/2} & - \quad 2^{5/4} 3^{-11/8} \delta & + \quad \mathcal{O}(\delta^2) \\
r_h^2 v_3 & \approx & \;\;\, 1/3^{1/2} & + \quad 2^{5/4} 3^{-11/8} \delta & + \quad \mathcal{O}(\delta^2) .
\end{array}
\end{equation}
where $\delta = [(b-b_c)/r_h]^{1/2}$. 
Applying the approximation 
\begin{equation}
K(k) \approx \ln \left[ \frac{4}{(1-k^2)^{1/2}} \right]
\end{equation}
to Eq.~(\ref{scat-ang-sol}), it is straightforward to show that
\begin{equation}
\theta(b) \approx -\frac{1}{2} \ln \left[ \frac{b - b_c}{ \beta_4 b_c}  \right]\,,
\end{equation}
where $\beta_4 = 108 e^{-2 \pi} \approx 0.201684$.

\bibliographystyle{aps}

\begin{thebibliography}{9}
\bibitem{footnote1}
A slightly more accurate approximation for the glory may be obtained 
by solving Eq.~(\ref{scat-ang-sol}) numerically to obtain 
$b_g = 1.6124635$ and $db/d\theta |_{\theta = \pi} = 0.00121996$ 
(c.f. $b_g = 1.6124620$ and $db/d\theta |_{\theta = \pi} = 0.00121416$ 
from logarithmic approximation~(\ref{strong-field-defl})) 
leading to alternative numerical constants $0.019930$ and $1.612463$ 
in Eq.~(\ref{glory-approx}).
\bibitem{footnote2}
Note that we have set the speed of sound equal to unity. 
In Eq.~(\ref{glory-approx}) the speed of sound should be reinserted
to obtain the numbers in usual kinematical units. Analogously, in
Eq.~(\ref{schw-glory-approx}) the speed of light should also be reinserted.
The glory peak will be 170 times weaker for the
canonical acoustic hole if 
$
 \omega_{\text{sound}}/c_{\text{sound}}=
 \omega_{\text{sf}}/c_{\text{light}},
$
for holes with the same horizon size. 
Here, $\omega_{\text{sf}}$ stands for the massless scalar field frequency.
\end{thebibliography}

\end{document}